\documentclass{article}

\usepackage{arxiv}
\usepackage{siunitx}

\usepackage[utf8]{inputenc} 
\usepackage[T1]{fontenc}    
\usepackage{hyperref}       
\usepackage{url}            
\usepackage{booktabs}       
\usepackage{amsfonts}       
\usepackage{amsmath}
\usepackage{nicefrac}       
\usepackage{microtype}      
\usepackage{lipsum}     
\usepackage{graphicx}
\usepackage{doi}
\usepackage{multirow}

\usepackage{xcolor}
\title{Meshless Monte Carlo Radiation Transfer Method for Curved Geometries using Signed Distance Functions}

\author{ \href{https://orcid.org/0000-0002-7725-5162}{\includegraphics[scale=0.06]{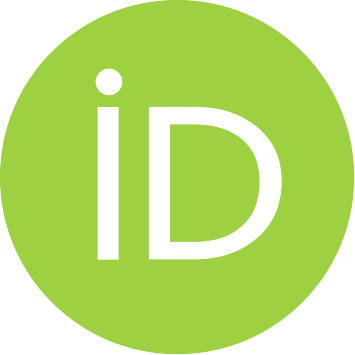}\hspace{1mm}Lewis McMillan}\\
    SUPA School of Physics and Astronomy\\
    University of St Andrews\\
    St Andrews, Scotland\\
    \texttt{lm959@st-andrews.ac.uk} \\
    \And
    \href{https://orcid.org/0000-0003-3403-0614}{\includegraphics[scale=0.06]{orcid.pdf}\hspace{1mm}Graham D.~Bruce} \\
    SUPA School of Physics and Astronomy\\
    University of St Andrews\\
    St Andrews,Scotland\\
    \texttt{gdb2@st-andrews.ac.uk} \\
    \And
    \href{https://orcid.org/0000-0001-6534-9009}{\includegraphics[scale=0.06]{orcid.pdf}\hspace{1mm}Kishan Dholakia} \\
    SUPA School of Physics and Astronomy\\
    University of St Andrews\\
    St Andrews, Scotland\\
    and\\
    Department of Physics, College of Science\\
    Yonsei University\\
    Seoul 03722, South Korea\\
    \texttt{kd1@st-andrews.ac.uk} \\
}

\renewcommand{\shorttitle}{\textit{signedMCRT}}

\begin{document}
\maketitle

\begin{abstract}

Significance:
Monte Carlo radiation transfer (MCRT) is the gold standard of modeling light transport in turbid media. 
Typical MCRT models use voxels or meshes to approximate experimental geometry.
A voxel based geometry does not allow for the accurate modeling of smooth curved surfaces, such as may be found in biological systems or food and drink packaging.

Aim:
We present our algorithm which we term signedMCRT (sMCRT), a new geometry-based method which uses signed distance functions (SDF) to represent the geometry of the model. SDFs are capable of modeling smooth curved surfaces accurately whilst also modeling complex geometries.

Approach:
We show that using SDFs to represent the problem's geometry is more accurate and can be faster than voxel based methods. sMCRT, can easily be incorporated into existing voxel based models.

Results:
sMCRT is validated against theoretical expressions, and other voxel based MCRT codes.
We show that sMCRT can accurately model arbitrary complex geometries such as microvascular vessel network using SDFs. In comparison to the current state-of-the-art in MCRT methods specifically for curved surfaces, sMCRT is up-to forty-five times more accurate. 

Conclusions:
sMCRT is a highly accurate, fast MCRT method that outperforms comparable voxel based models due to its ability to model smooth curved surfaces. sMCRT is up-to three times faster than a voxel model for equivalent scenarios. sMCRT is publicly available at \url{https://github.com/lewisfish/signedMCRT}.
\end{abstract}

\keywords{Monte Carlo, Light Transport, signed distance functions, SDF, geometry, meshless}



\section{Introduction}


The modeling of light transport is important to our understanding of how light interacts with turbid media. 
It allows us to make predictions of the viability of treatment modalities~\cite{campbell2015monte,barnard2020could}, simulate the behavior of complex shaped light in highly scattering media~\cite{mcmillan2021imaging}, retrieve images of objects in highly scattering media~\cite{lyons2019computational} and optimize light sensors in the food and drink industry~\cite{qin2009monte} among other applications.
The current "gold standard" of modeling light transport in turbid media is the Monte Carlo radiative transfer (MCRT) method.
MCRT can model light transport in arbitrary 3D geometries and model several micro-physics phenomena such as Raman scattering~\cite{keller2010monte,wang2014monte}, fluorescence~\cite{liu2003experimental,welch1997propagation}, polarization~\cite{reidt2016polarised,ramella2005one,ramella2005two}, and has been applied to problems ranging from light propagation in dynamic fluid systems~\cite{vandenbroucke2018monte,harries2019torus} to simulating thermal gradients in illuminated tissue~\cite{mcmillan2021development,jeynes2019monte}.

In order to simulate the transport of light through a medium, the geometry of the problem must modeled.
Most Monte Carlo codes rely on voxels~\cite{wang1995mcml,marti2018mcmatlab,Boas:02}, or meshes~\cite{fang2010mesh,wilson2011mesh} to approximate the geometry of the problem.
Voxel models are only suitable for the simplest problems that do not require accurate treatment of curved surfaces, due to their cubic nature~\cite{binzoni2008light}.
Curved surfaces arise in many problems where MCRT maybe used, for example the propagation of light through: an optical system, the anatomy of animals or humans such as the brain or vascular networks, among many other possible examples. 
In contrast, mesh based models can accurately treat curved surfaces but usually require specialist software to create high quality meshes which can also be computationally expensive to produce, and MCRT codes require extensive software engineering to incorporate meshes in a computationally efficient manner.

A number of previous methods have been introduced to tackle the problem of smooth surfaces in voxel based models.
Tran and Jacques preprocessed the voxels to determine where the material interfaces are, and computed surface normals for each voxel which can then be smoothed via interpolation to create curved surfaces~\cite{tran2020modeling}.
Whilst this method, on the whole, accurately models curved surfaces, it can have an increased memory footprint, and is more computationally intensive.
Alternatively, implicit surfaces can be defined using mathematical formulae~\cite{periyasamy2014monte,majaron2015monte}.
However, while this method of defining mathematical surfaces allows the modeling of smooth surfaces, it has the drawback that each surface needs an accompanying intersection and surface normal routine which can be computationally costly to evaluate and increases the workload on the programmer.

In this work, we present a novel Monte Carlo radiative transfer model where we eschew the common voxel or mesh based approaches for an approach based upon signed distance functions (SDFs), which we call signedMCRT (sMCRT)\@.
SDFs have been commonly used to define implicit surfaces in computer graphics~\cite{hart1989ray}, video games~\cite{evans2015learning}, and computer vision~\cite{park2019deepsdf}. Recently, there has been considerable interest in using neural networks to define SDFs from point clouds and meshes. This interest has been led by computer graphics and deep learning researchers, looking for memory efficient representations of meshes and/or point clouds at high spatial resolutions~\cite{park2019deepsdf,sitzmann2020implicit,takikawa2021neural}.

SDFs allow the efficient transport of photon packets through the modeled geometry using sphere tracing, which is faster, in most cases, compared to traditional ray tracing methods used in MCRT~\cite{hart1989ray}.
SDFs, while being similar to the approach of mathematical surfaces, do not need individual intersection routines as they are naturally included in the SDF definition. Moreover, we can use numerical differentiation to provide the surface normals.
SDFs also require little effort to incorporate into existing voxel based codes, requiring tweaks to the optical depth integration and geometry initialization routines.
These features of SDFs make them an attractive alternative to voxel models.

We show in this paper that sMCRT is more malleable and accurate the traditional voxel models for curved surfaces and more accurate than current solutions (by up-to 45 times more accurate) to the voxel discretization problem. Moreover, we demonstrate that sMCRT is faster (for certain problems).
MCRT can be faster, or slower than traditional voxel based methods depending on the output required. When MCRT is used to retrieve the intensity at each point of a volume of interest, there is fractional increase in runtime of sMCRT (up-to 0.6 times slower) when compared with a traditional voxel based code. However, for problems where the desired output is only the light distribution at a single plane of interest, then sMCRT is fractionally faster than the traditional voxel-based methods (up-to 3 times faster).

\section{Methods}
\subsection{Monte Carlo Radiation Transfer Algorithm}

The radiation transfer equation (RTE) describes the transfer of energy in a medium. However, analytical solutions for the RTE only exist for simple geometries.
Therefore, numerical methods such as the diffusion method~\cite{star1995diffusion} or the Monte Carlo radiation transfer method (MCRT) must be used to compute a solution.

MCRT uses interaction probabilities and probability distribution functions that describe the physics of light transport, to model light transport though turbid and non-turbid media.
Each photon is propagated a distance $\tau/\mu_t$, where $\tau$ is the optical distance [-] and $\mu_t$ [\si{\per\centi\metre}] is the extinction coefficient, before it interacts with the medium.
The value of $\tau$ is sampled from the probability distribution function for the mean free path of a photon using the Monte Carlo method~\cite{wang1995mcml}, as shown in Equation~\ref{eqn:taurand}, where $\xi$ is a random number drawn in the range [0, 1]. For voxel models the extinction coefficient can differ from voxel to voxel.
MCRT is highly accurate (as the number of photons simulated tends to infinity) and can be used to model the light transport in arbitrary 3D geometries provided these geometries can be modeled.

\begin{equation}
    \tau = -log(\xi)
    \label{eqn:taurand}
\end{equation}

The MCRT code presented in this work is broadly based upon previous MCRT codes used in various astronomical, medical, and bio-photonics applications~\cite{mcmillan2021imaging,wood1999model,barnard2018quantifying,finlaysondepth}.
We use the same routines for releasing photons, input/output, scattering, random number generation, and helper routines.
What differs in this work is the optical depth integration routine, and the geometry modeling method, which are accomplished by the use of signed distance functions.

\subsection{Signed Distance Functions}

Signed distance functions determine the distance from a point $p$, to the boundary of a specified shape.
The function returns a positive value if $p$ is outside the boundary, and a negative value if inside the boundary.
Formally, this can be described using level set representation.
In level set representation contours are modeled at the zero-level set ($\phi=0$) of a function defined in a higher dimension.
Let $\Phi:\Omega \rightarrow \mathcal{R} ^3$ be a Lipchitz function that refers to a level set representation for a given shape $\mathcal{S}$~\cite{paragios2002matching} then:
\begin{equation}
    \Phi_s(x,y,z)=\begin{cases}
    0, \qquad \qquad \qquad \, \qquad (x,y,z)\in \mathcal{S}\\
+D((x,y,z), \mathcal{S})>0, (x,y,z)\in \mathcal{R}_\mathcal{S}\\
    -D((x,y,z), \mathcal{S})<0, (x,y,z) \in [\Omega- \mathcal{R}_\mathcal{S}]
    \end{cases}
\end{equation}

An example of an SDF is shown in Equation~\ref{eqn:sphere} for a sphere, where $r$ is the radius of the sphere, and $p$ is the position of a photon:

\begin{align}
    D_{sphere}(x,y,z) &= \left|p\right| - r \\
    p &= [x,y,z]
    \label{eqn:sphere}
\end{align}

\begin{figure}[!htbp]
    \centering
    \includegraphics[width=0.75\textwidth]{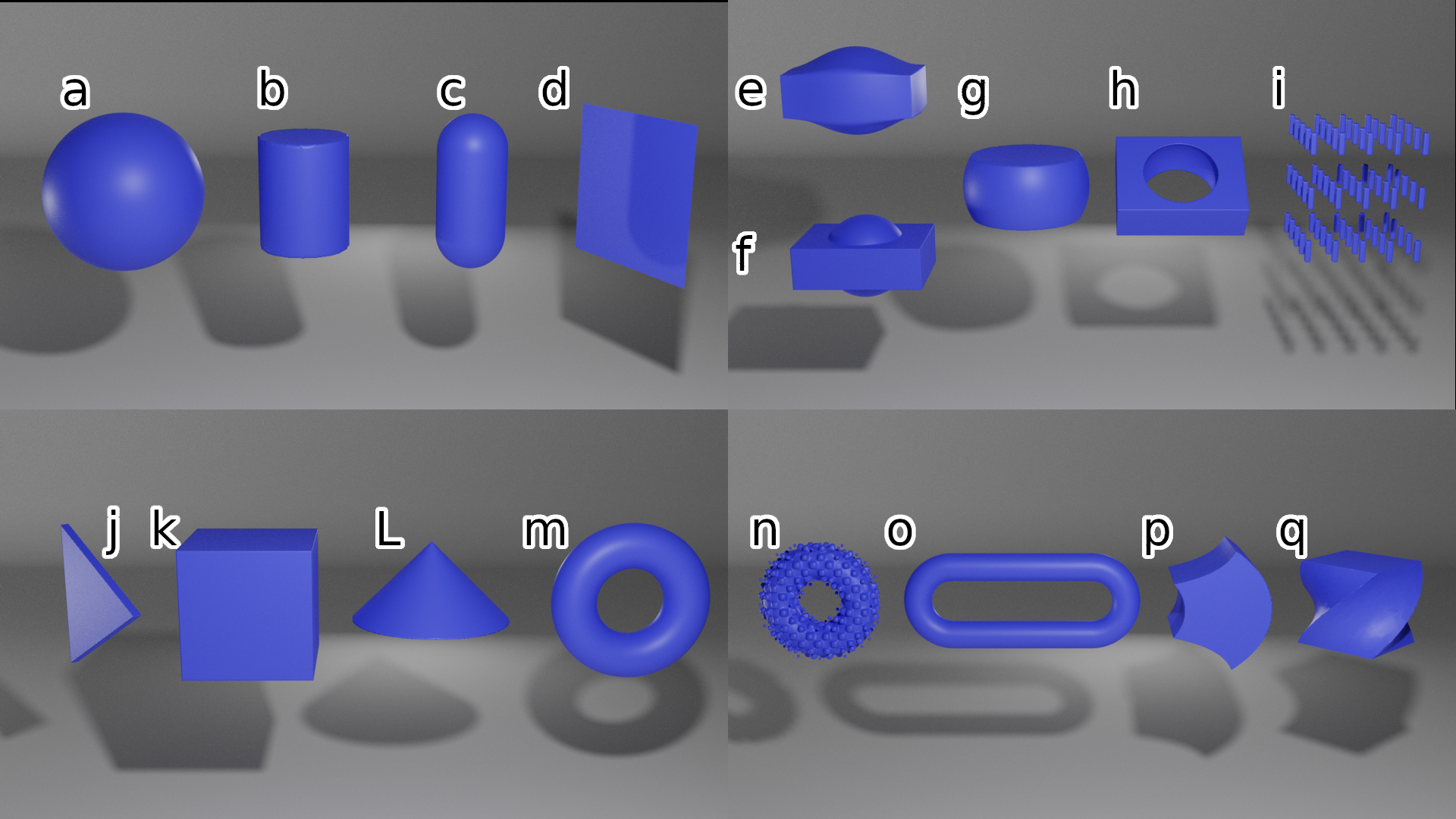}
    \caption{Several examples surfaces that can be created by SDFs, rendered in Blender. For illustrative purposes, SDFs are voxelized in sMCRT then transformed to a mesh using Skimage's~\cite{van2014scikit} marching cubes algorithm and then rendered using Blender~\cite{blender}. The left two panels show a subset of basic shapes calculated using SDFs (a to d and j to m). The right two panels show a subset of possible operations on SDFs: smooth (e) and non-smooth union (f), intersection (g), subtraction (h), repetition (i), displacement (n), elongation (o), bend (p), and twist (q).}
    \label{fig:gallery}
\end{figure}

SDFs can easily be translated, rotated, twisted and scaled among many other operations.
Constructive solid geometry (CSG) operations such as union, intersection and difference can also be used on the SDFs.
Figure~\ref{fig:gallery} shows a subset of shapes, and possible operations on SDFs~\cite{hart1996sphere,quilez}.

\subsection{MCRT SDF Algorithm}

\begin{figure}[!htbp]
    \centering
    \includegraphics[width=0.8\textwidth]{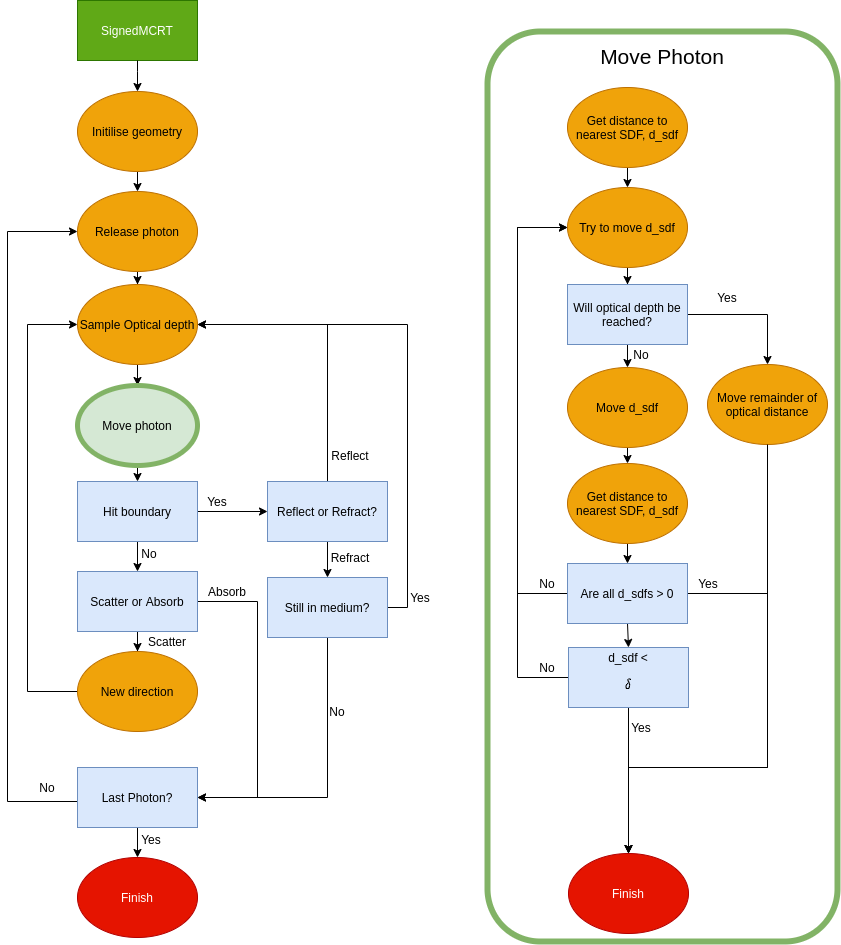}
    \caption{Flow diagram of an MCRT code (left). Right panel shows the additional steps needed to incorporate signed distance functions into the optical depth integration routine which governs the movement of photon packets through the simulated media.}
    \label{fig:algo}
\end{figure}

To incorporate SDFs into a pre-existing voxel based MCRT code, requires only relativity small adjustments: modifications to the geometry initialization routine, and to the optical depth integration routine. An overview of the complete MCRT algorithm is shown in the left panel of Figure~\ref{fig:algo}.

To create the geometry in voxel based models, each voxel is independently assigned a set of optical properties (scattering and absorption coefficients, refractive index, and anisotropy $g$ value).
In sMCRT the geometry is initialized by selecting the functional form, size, and location of SDF(s) required to model the problem, applying any CSG operations required to generate more complex shapes, and finally setting the optical properties for each SDF\@.
Each SDF has its own set of optical properties, which include scattering and absorption coefficient, refractive index and the anisotropy $g$ value. We then create a bounding box around all the SDFs, which gives us a simulation volume of interest.

In voxel based MCRT codes, each photon packet is randomly ascribed a specific optical path length that it travels before an interaction, such as scattering or absorption, according to Equation~\ref{eqn:taurand} and is scaled by $\mu_t$ ($\mu_t$ can be different for each voxel).
The photon packet is then propagated through the voxel grid using ray tracing until it reaches that interaction point or leaves the voxel grid.

\begin{figure}[!htbp]
    \centering
    \includegraphics[width=0.5\textwidth]{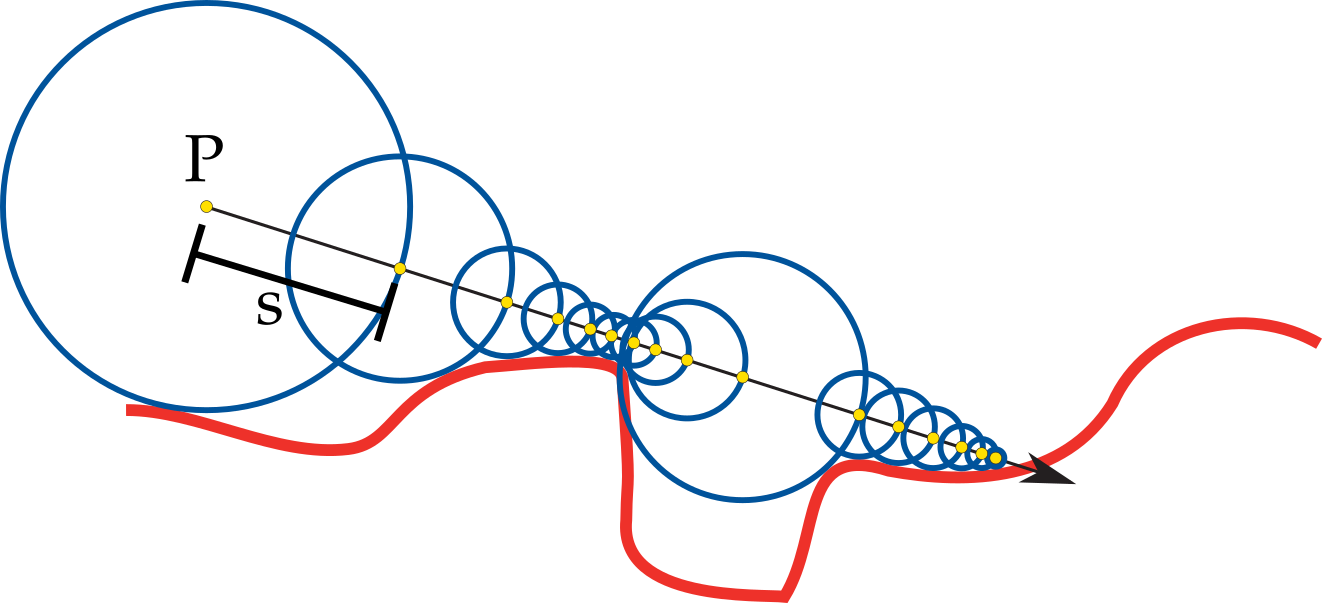}
    \caption{Example of sphere tracing. Starting a position $P$, the photon is propagated by a step size, $s$ (represented here by the blue circle), equal to the distance to the nearest surface (red line) until the step size is under some threshold $\delta$.}
    \label{fig:sphere}
\end{figure}

In our SDF based MCRT algorithm, the first step in the SDF optical depth integration routine is the same as in the voxel case, i.e. randomly assign an optical depth.
As before, this is calculated using Equation~\ref{eqn:taurand} and $\mu_t$ can be different for each SDF\@.
The next step is to acquire the distance from the current position of the photon packet to the nearest boundary.
This is computed by using the SDFs to 
calculate the distance to each boundary in the modeled geometry, and taking the minimum value ($d_{sdf}$).
This process is called sphere tracing~\cite{hart1996sphere}, and is illustrated in Figure~\ref{fig:sphere}.

If the remaining optical depth for the photon packet is less than $d_{sdf}$, the photon packet undergoes some interaction, and the optical depth integration routine restarts.
If the optical depth is not reached, then we move the full distance $d_{sdf}$, and then recalculate the distances to all boundaries.
If the SDF of the bounding box returns a positive value we are outside the volume of interest, so we terminate the packet and start a new packet.

If the SDF for the bounding box returns a negative value, we then check if the smallest distance, $d_{sdf}$, is less than some threshold, $\delta$. In this case, the photon packet is on a boundary so we need to check if there is a change in refractive index.
If there is a change in refractive index we calculate the Fresnel coefficients and the surface normals, then reflect or refract the photon packet.
If $d_{sdf}$ is larger than $\delta$, and all distances to the SDFs are not positive then we start this whole process again until one of the exit conditions has been met.
The above algorithm is shown in the right panel of Figure~\ref{fig:algo}.

\section{Results and Discussion}
\subsection{Validation}

To ensure that our novel SDF-based geometry method works accurately, we validate our algorithm against a theoretical expression, and two other MCRT codes.
All simulations are fully parallelized with OpenMP and were run on a workstation with an AMD Ryzen 9 3950X 16-Core Processor with 64~GB RAM utilizing the full 32 threads available.

We first compare sMCRT's accuracy by computing the average number of scattering events occurring to a photon in an isotropic sphere~\cite{rybicki1991radiative}.

For a photon's random walk from the center to the edge of a uniformly scattering sphere of radius $r$, the average number of scatterings that take place can be written as (see Supporting Information):
\begin{equation}
    N\approx\frac{\tau^2}{2}+\tau
    \label{eqn:scateqn}
\end{equation}

\begin{figure}[!htbp]
    \centering
    \includegraphics[width=0.75\textwidth]{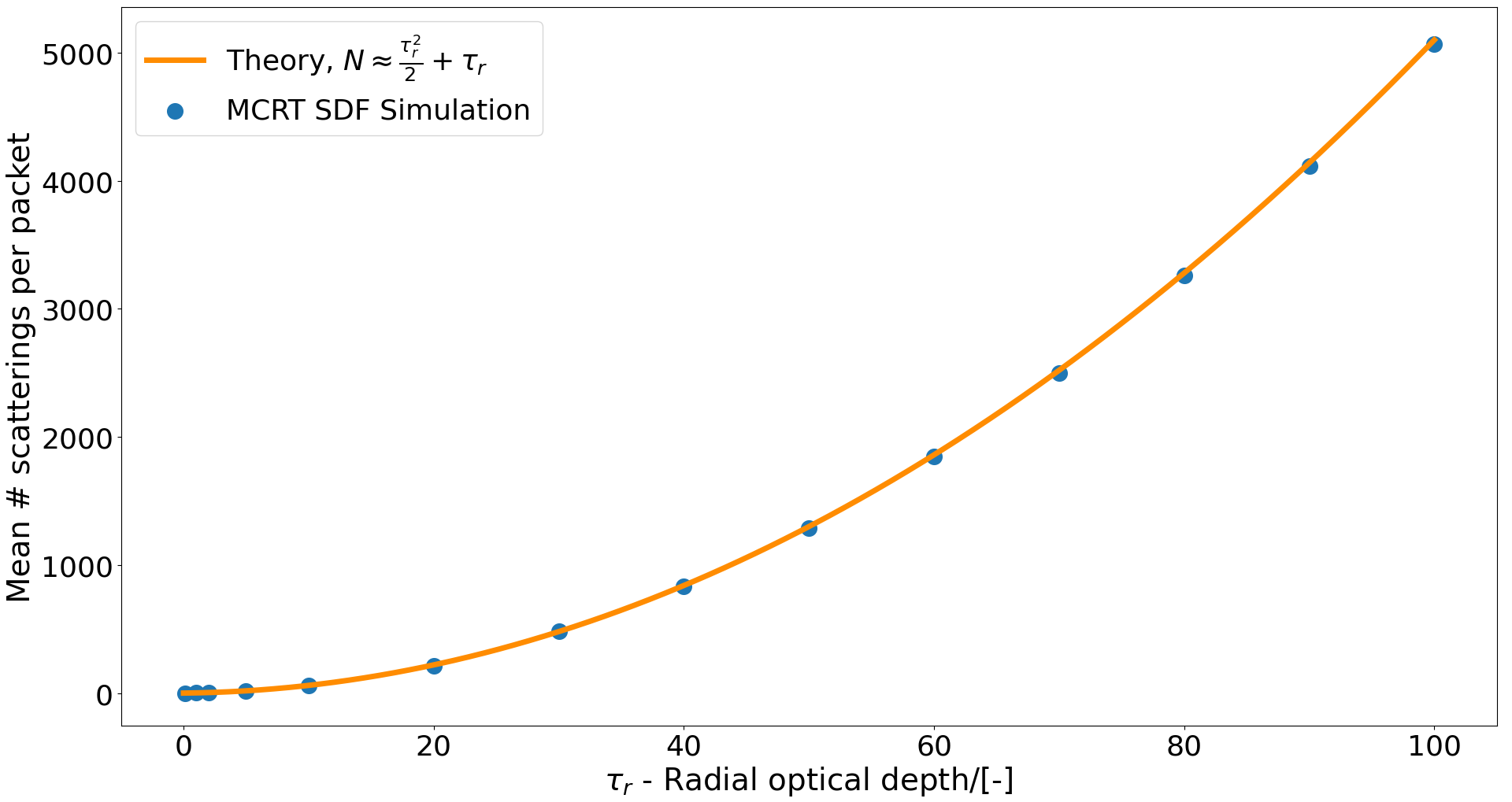}
    \caption{Agreement of analytical expression (Equation~\ref{eqn:scateqn}) and sMCRT for several optical depths in the range [0.01, 100]. Photons are released from the center of an isotropic scattering sphere. The optical density (scattering coefficient) is varied and the average number of scattering per photon packet is recorded.}
    \label{fig:tauexp}
\end{figure}

To compare Equation~\ref{eqn:scateqn} to sMCRT, we model a sphere of radius 0.5~\si{\centi\metre}, vary the radial optical depth between 0.1 and 100, and release 10 million photons isotropically from its center.
The agreement of the code and analytical expression can be seen in Figure~\ref{fig:tauexp}.

We also validated sMCRT against our previous voxel based MCRT code~\cite{mcmillan2021imaging,mcmillan2021development,mcmillan2020advanced}, and S. Jacques \textit{et al.} MCRT code~\cite{jacques1993photobleaching}. We validate against S. Jacques code as it incorporates all the relevant physics we need in an MCRT code; scattering, absorption, and refractive index mismatches.

For this validation, the medium is set up as a semi-infinite slab and light is uniformly incident on the surface of the slab (negative z direction) and propagates until it is absorbed or escapes via the top surface (positive z direction).
We then fit against Equation~\ref{eqn:fluedepth} to compare between codes.
\begin{equation}
    \Psi(z)=\Psi_0(C_1e^{-(zk_1 / \delta)} - C_2e^{-(zk_2/\delta)})
    \label{eqn:fluedepth}
\end{equation}

Here $\Psi(z)$ is the penetration of the incident light or equivalently the fluence rate [\si{\watt \per\square\centi\metre}], $\Psi_0$ is a normalization constant [\si{\watt \per\square\centi\metre}], $C_n$ and $k_n$ are fitted coefficients [-], and $\delta$ is the optical penetration depth, defined as $\delta=1/\sqrt{3\mu_a(\mu_a+\mu_s(1-g))}$ [\si{\centi\metre}].
The optical properties for the slab are shown in Table~\ref{tab:jacqprops}, where we use the Henyey-Greenstein phase function~\cite{henyey1941diffuse} with a $g$ of 0.9, and we model two wavelengths in separate simulations.
The refractive index for the medium was set to 1.38 to mimic the rat skin used in S. Jacques code~\cite{jacques1993photobleaching}, and for the surrounding medium a refractive index of 1.0 was set, to mimic air.

\begin{table}[!ht]
\caption{Table of optical properties and determined coefficients from Jacques \textit{et al.}~\cite{jacques1993photobleaching}.}
\centering
\begin{tabular}{llllllll}
                                   & \multicolumn{1}{c}{{ Absorption}} & \multicolumn{1}{c}{{ Scattering}}    & \multicolumn{4}{c}{{ Penetration}}          &             \\
\multicolumn{1}{l|}{Wavelength [\si{\nano\metre}]} & \multicolumn{1}{l|}{$\mu_a$ [\si{\per\centi\metre}]} & \multicolumn{1}{l|}{$\mu_s(1-g) [\si{\per\centi\metre}]$} & C1   & k1   & C2   & \multicolumn{1}{l|}{k2}   & $\delta [\si{\per\centi\metre}]$ \\ \hline
\multicolumn{1}{l|}{420}           & \multicolumn{1}{l|}{1.8}             & \multicolumn{1}{l|}{82}                 & 5.76 & 1.00 & 1.31 & \multicolumn{1}{l|}{10.2} & 0.047       \\
\multicolumn{1}{l|}{630}           & \multicolumn{1}{l|}{0.23}            & \multicolumn{1}{l|}{21}                 & 6.27 & 1.00 & 1.18 & \multicolumn{1}{l|}{14.4} & 0.261      
\end{tabular}
\label{tab:jacqprops}
\end{table}

\begin{figure}[!htbp]
    \centering
    \includegraphics[width=0.75\textwidth]{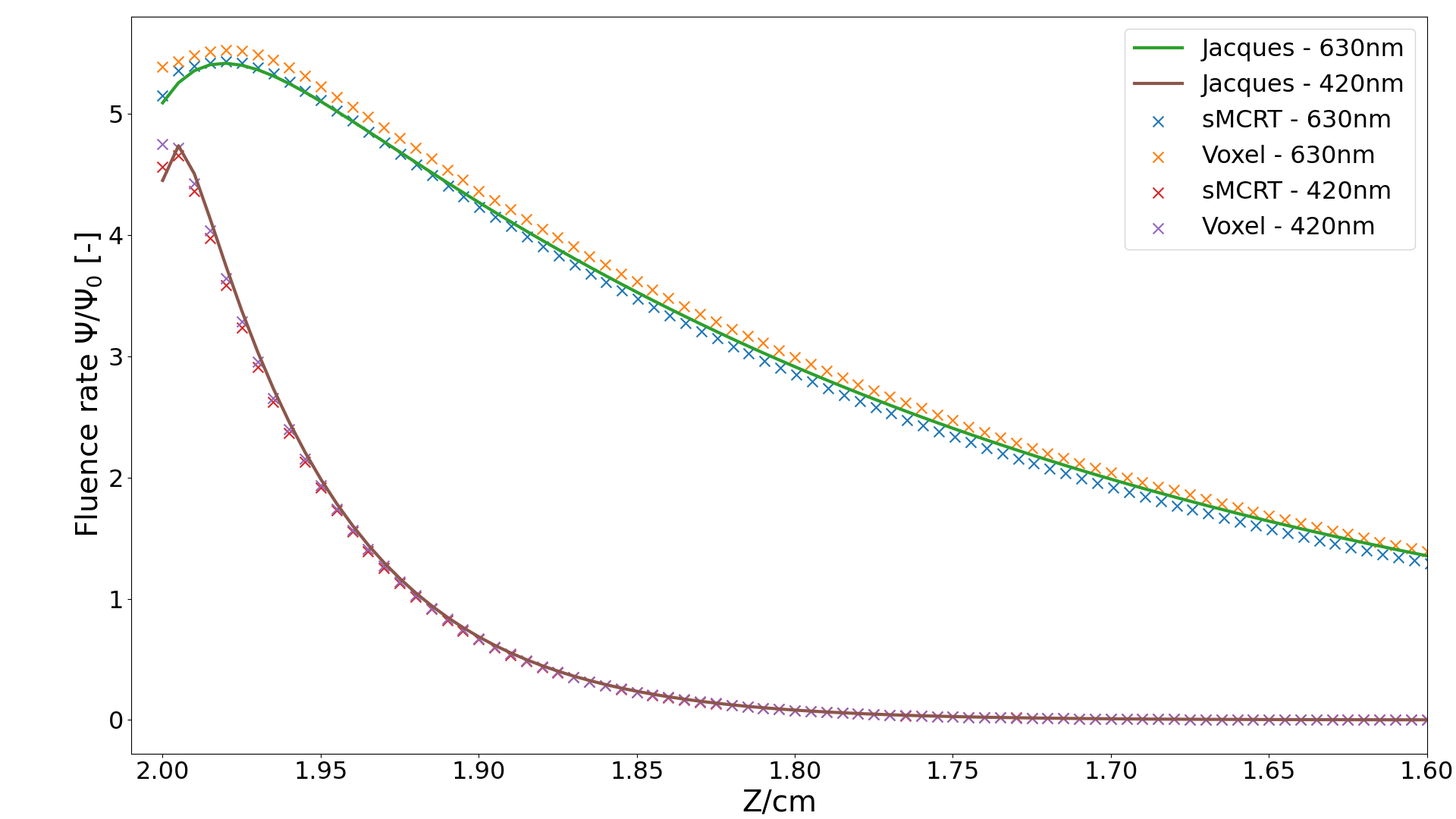}
    \caption{Validation of sMCRT against our previous voxel based model, and S. Jacques \textit{et al.} MCRT model. The simulation medium is a semi-infinite slab (infinite in the $x$ and $y$ dimensions), and has the optical properties as in Table~\ref{tab:jacqprops}. The medium is uniformly illuminated via the top surface, i.e. is incident from the left of this figure.}
    \label{fig:jacquesfit}
\end{figure}

As evidenced in Figure~\ref{fig:jacquesfit}, sMCRT gives a better match to the results in S. Jacques \textit{et al.} work than our previous voxel model.
An exact match is not possible, due to the difference in the code underlying workings such as cylindrical fluence bin shape used by S. Jacques \textit{et al.} versus our rectangular bin shape.
sMCRT is more accurate than the voxel model at 420~\si{\nano\metre} as it incorporates the left hand-side of the peak near the surface whereas the voxel model levels off after the peak fluence value.

\subsection{Comparing Voxel Models and sMCRT}

In this section we compare the accuracy and speed of a voxel based MCRT code to that of our new SDF based MCRT code, sMCRT.
We devise two tests based upon speed and accuracy.

\begin{figure}[!htbp]
    \centering
    \includegraphics[width=.75\textwidth]{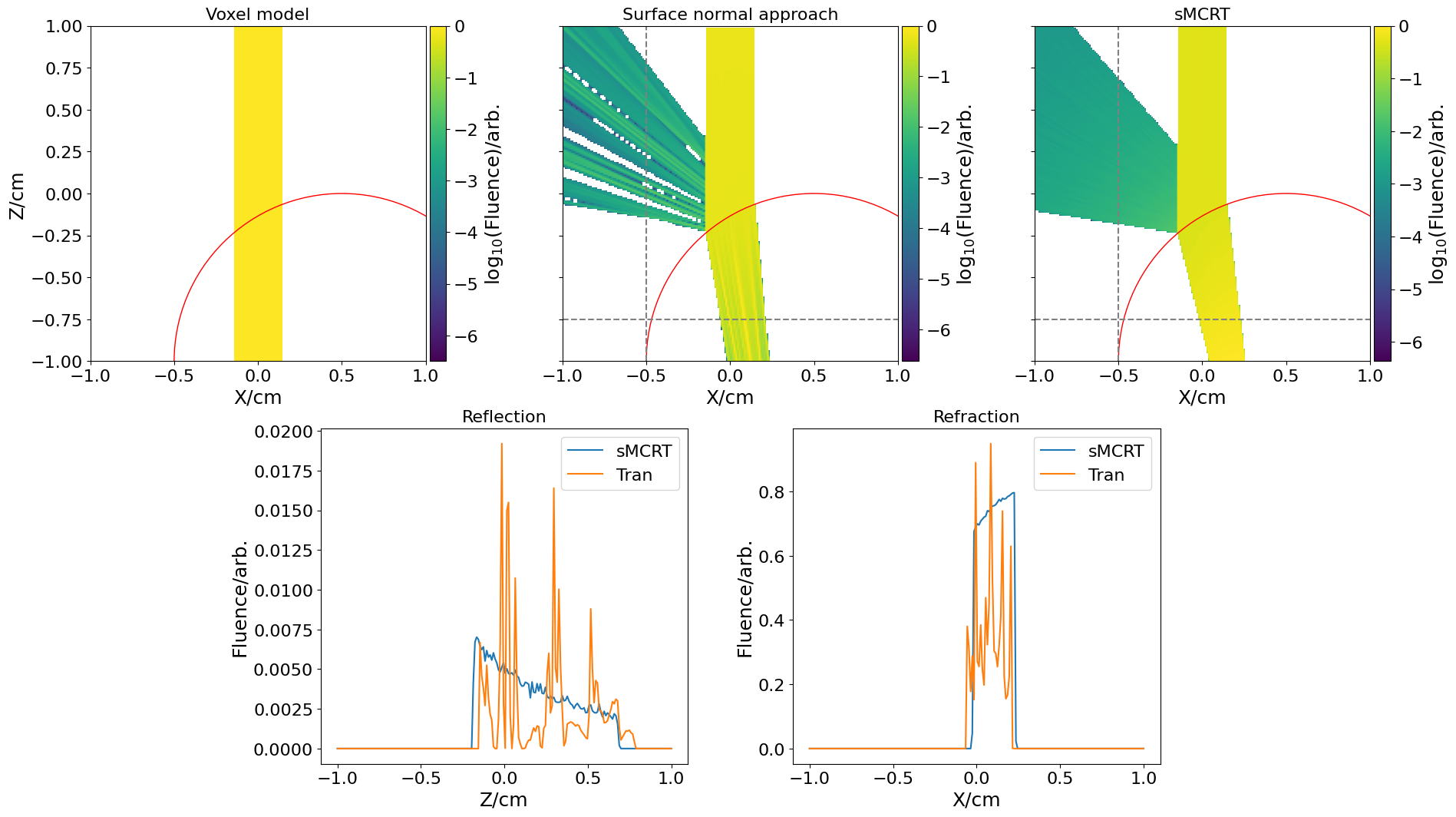}
    \caption{Comparison of simulation accuracy between voxel, A.P Tran and S. Jacques surface normal approach~\cite{tran2020modeling} interpolation method and sMCRT models. A glass sphere (red circle) with refractive index 1.46 and radius 1.0~\si{\centi\metre} is placed at the (0.5, 0.0, -1.0)~\si{\centi\metre} of the simulated medium. Light is incident on the top surface of the sphere with a beam radius of 0.3~\si{\centi\metre}. Figure shows a slice of fluence though the center of the medium for the voxel model (top-left), surface normal approach (top-middle), sMCRT (top-right), and the difference between sMCRT and the surface normal approach along the gray lines in top right two panels (bottom-row). This clearly shows that the voxel geometry cannot model curved surfaces and their reflections and refractions. The surface normal approach allows semi-accurate modeling, whereas the sMCRT model can model these reflections and refractions with greater accuracy.}
    \label{fig:curvetest}
\end{figure}

The first test is a simple test of modeling a smooth surface: for this we use the test case in Tran and Jacques surface normal approach~\cite{tran2020modeling}.
They model a glass sphere (refractive index 1.46) and let a beam of light, with a waist of 0.3~\si{\centi\metre}, be incident on the sphere (in the negative z direction). 
Figure~\ref{fig:curvetest} shows a slice of the fluence through the sphere for sMCRT, a voxel based MCRT code, and Tran and Jacques code.

Here the voxel model (top right panel) does not exhibit the expected reflection and refraction of light due to the lack of smooth surfaces in the modeled geometry.
The sphere is discretized onto a voxel grid which only has flat surfaces, leading to the lack of accurate refractions/reflections.
In comparison sMCRT model (top left panel) accurately replicates the refractions/reflections at the glass/air boundary. 

The surface normal approach (bottom left panel) shows an improvement on the basic voxel model alone, it still suffers from inaccuracies.
These inaccuracies, as evidenced from the missing reflections and refractions, arise from their model still being based upon voxels.
They interpolate the surface normals at the refractive index mismatches, however this new virtual surface is not in the correct place for accurate reflections and refractions due to discretization errors.

The difference in accuracy between the surface normal approach and sMCRT can be seen in the lower row of Figure \ref{fig:curvetest}, which shows a line profile across each of the reflected and refracted beams for the two approaches.
The beam profiles produced by sMCRT exhibit a reduction in the root mean square deviation from the expected beam profile by a factor of 10 for the reflected beam and by a factor of 45 for the refracted beam, when compared to the surface normal approach.

sMCRT also outperforms the surface normal approach on speed for this test case. sMCRT took on average 61 seconds to run $\sim$ 27 million
photons compared to 250 seconds for the surface normal approach.
The speedup factor here is 4. Both codes were run on the full 32 threads available on our workstation.

\begin{figure}[!htbp]
    \centering
    \includegraphics[width=0.65\textwidth]{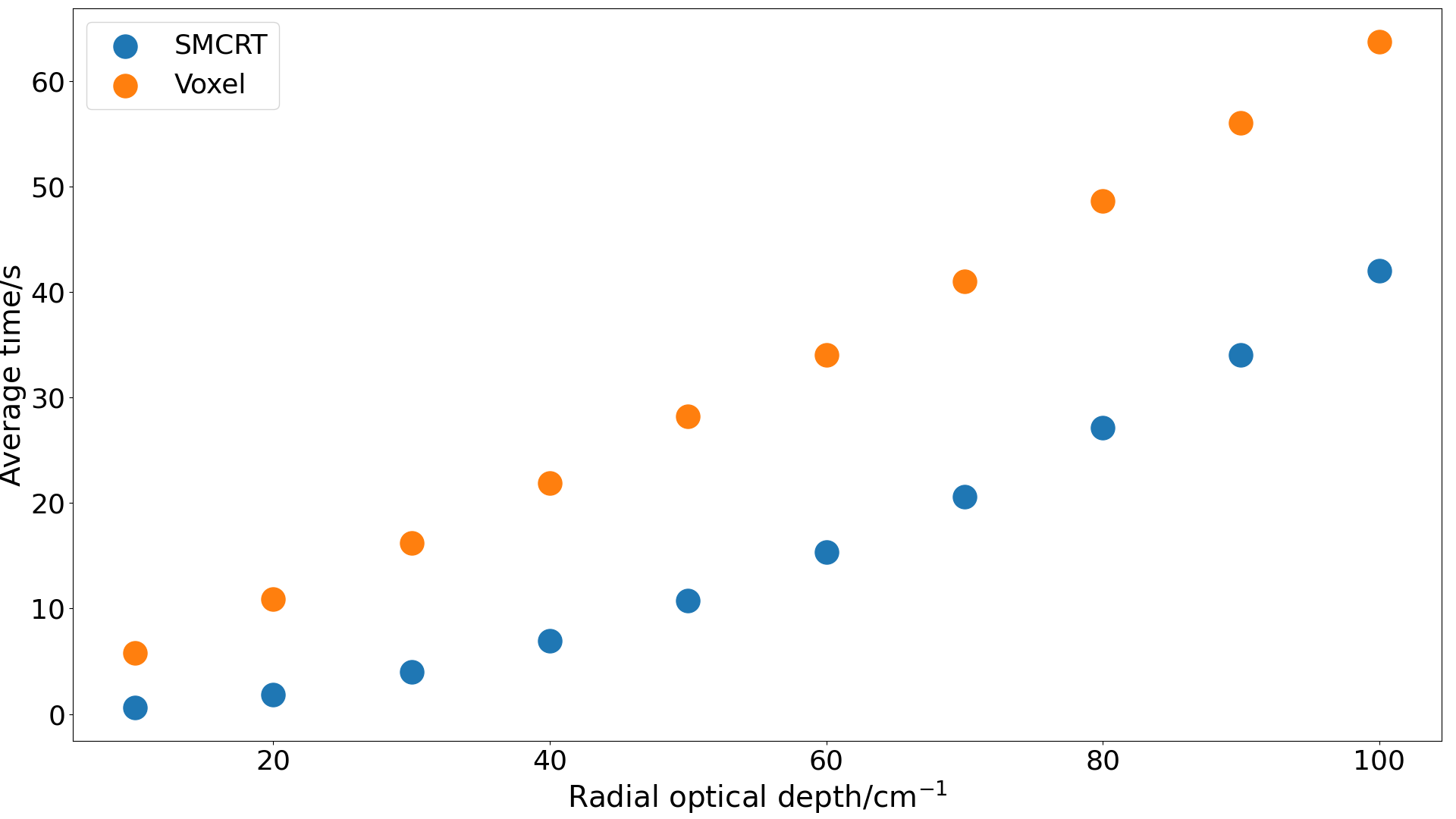}
    \caption{Comparison of speed between voxel and SDF based MCRT models. The time taken for the calculation of 10 million photon trajectories was recorded for a variety of optical depths.}
    \label{fig:speedtest}
\end{figure}

The second test is a comparison of speed between the SDF and voxel models.
Here we use the same setup from the validation section, an isotropic scattering sphere which releases photons isotropically from its center.
We vary the scattering coefficient and time the simulation for 10 million photons.
As shown in Figure~\ref{fig:speedtest}, the SDF model is up to 3 times faster when run on the same workstation.

This however may not be representative of the true speed difference between the SDF and voxel models in real use scenarios.
In another test, we calculated the average number of scatterings per photon, which requires no fluence computation and thus no grid to record fluence.
The most common use of MCRT codes is to compute the fluence throughout the simulated medium. The voxel model provides this with little extra computation.
However, sMCRT needs no such grid, so the introduction of a fluence counting grid will slow down the simulation.
We therefore, test the two models whilst computing the fluence for two different scenarios: the S. Jacques validations case, and the aforementioned isotropic sphere test.
Table~\ref{tab:speeddown} summarizes the results of these tests.
For the S. Jacques \textit{et al.} validation case, we use the same geometry as before and a fluence grid of size 200 $\times$ 200 $\times$ 400 voxels for both the SDF and voxel models. We run 10 million photons and use 32 threads on our workstation for both models.
The isotropic sphere test was run for 1 million photons and a grid of $200^3$ voxels for both the SDF and voxel models. The radial optical depth of the sphere was set to 100.



\begin{table}[!htb]
\caption{Timings of the SDF and voxel models for the S. Jacques \textit{et al.} validation case and the isotropic sphere test. Time is the average of three runs using the same seed, and the speedup is relative to the voxel model case for both tests.}
\centering
\begin{tabular}{c|c|c|l}
Test                                                                             & Model & Time [\si{\second}]                     & Speedup \\ \hline
\multirow{2}{*}{\begin{tabular}[c]{@{}c@{}}S. Jacques\\ Validation\end{tabular}} & sMCRT   & 33.73                      & 0.60     \\
                                                                                 & Voxel & 20.37                      & 1.0      \\ \hline
\multirow{2}{*}{\begin{tabular}[c]{@{}c@{}}Isotropic\\ sphere\end{tabular}}      & sMCRT   & \multicolumn{1}{l|}{75.76} & 0.73     \\
                                                                                 & Voxel & \multicolumn{1}{l|}{55.59} & 1.0     
\end{tabular}
\label{tab:speeddown}
\end{table}

The results show that the sMCRT is slower than the voxel model for both test cases by a factor of $\approx 1.5$.
It should be noted that sMCRT is a new code, whereas the voxel model has had several years of development and optimizations applied to it, meaning that sMCRT could reach the same level of performance as the voxel model with a similar effort applied.
sMCRT does however, give more accurate results for both of these problems at the expense of speed.

\section{Complex Geometry}

\begin{figure}[!htbp]
    \centering
    \includegraphics[width=0.8\textwidth]{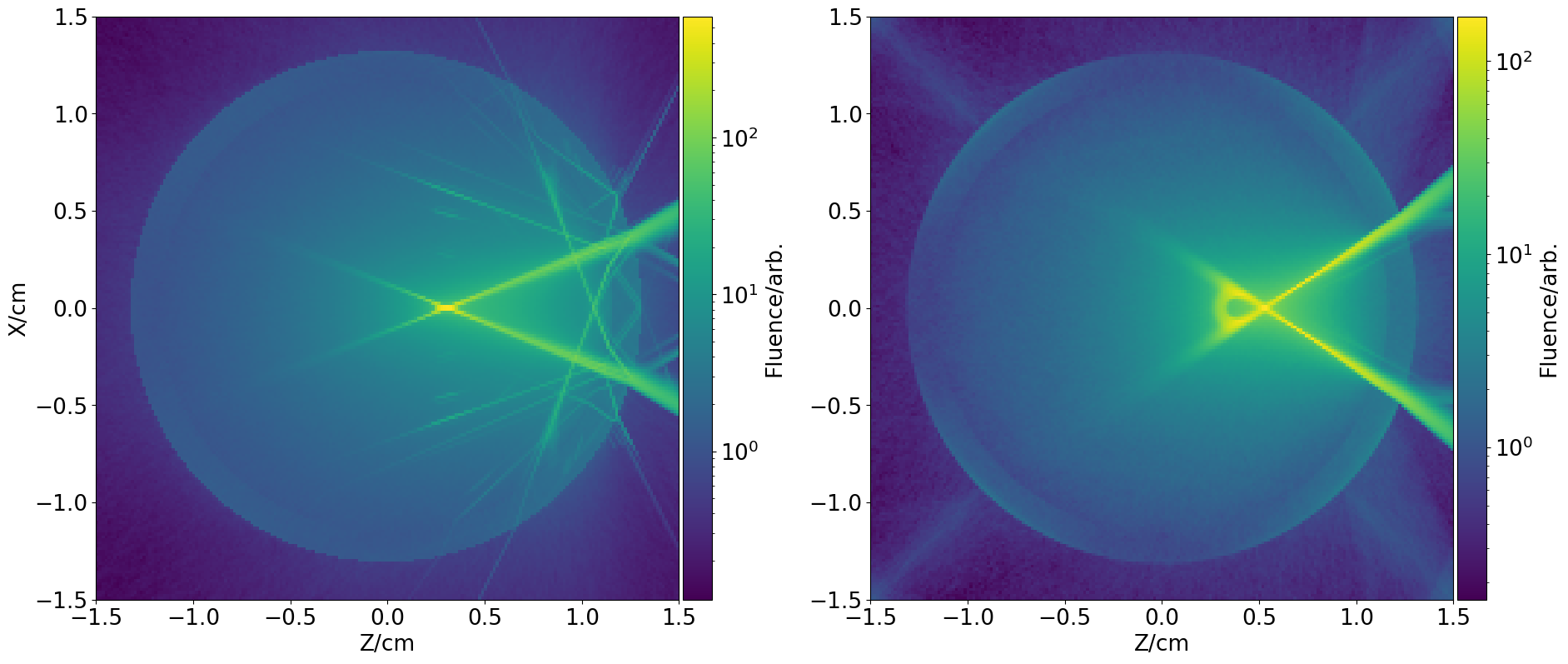}
    \caption{Comparison of fluence for the voxel model and sMCRT model in a glass bottle. Left panel shows the voxel model. Right panel shows the sMCRT. Both panels show the cross section of the bottle. This shows clearly that the voxel model cannot accurately model reflections/refraction in an experiment with curved surfaces, as it shows discretized reflections and refractions, whilst sMCRT shows the expected continuum of reflections and refractions.  For this example the optical properties are set for the contents $\mu_s$=2.5 \si{\per\centi\metre}, $\mu_a$=0.01 \si{\per\centi\metre}. The glass has no scattering and has the same absorption coefficient. The refractive index of the glass is 1.5 and 1.3 for the contents. Both the glass and contents of the bottle have a g value of 0.7. The bottles radius is 1.75~\si{\centi\metre}, and the glasses thickness is 0.2~\si{\centi\metre}}
    \label{fig:bottle}
\end{figure}

SDFs can be used in place of current voxel based geometries as they offer increased accuracy in the modeling of complex geometries, as well as the accurate treatment of smooth surfaces. 
We have recently used sMCRT in our simulations of Raman spectroscopy of alcoholic beverages through glass bottles~\cite{georgie2021}.
Accurate modeling of a bottle in a voxel geometry was not possible, as this did not allow accurate reflection/refractions at the glass/air and glass/alcohol interfaces. The lack of these accurate reflections meant a massive decrease in the accuracy of the simulation of the light distribution in the bottle, as the bottle acts as a cylindrical lens on the light, see SI~\ref{fig:bottle}.

Thus far we have shown that sMCRT can model simple geometries, so to demonstrate that sMCRT can model complex shapes, we model a blood vessel network embedded in tissue, and show that sMCRT can model arbitrary SDF generated neural networks, such as DeepSDF or SIREN, see SI Figure 1.

\begin{figure}[!htbp]
    \centering
    \includegraphics[width=0.8\textwidth]{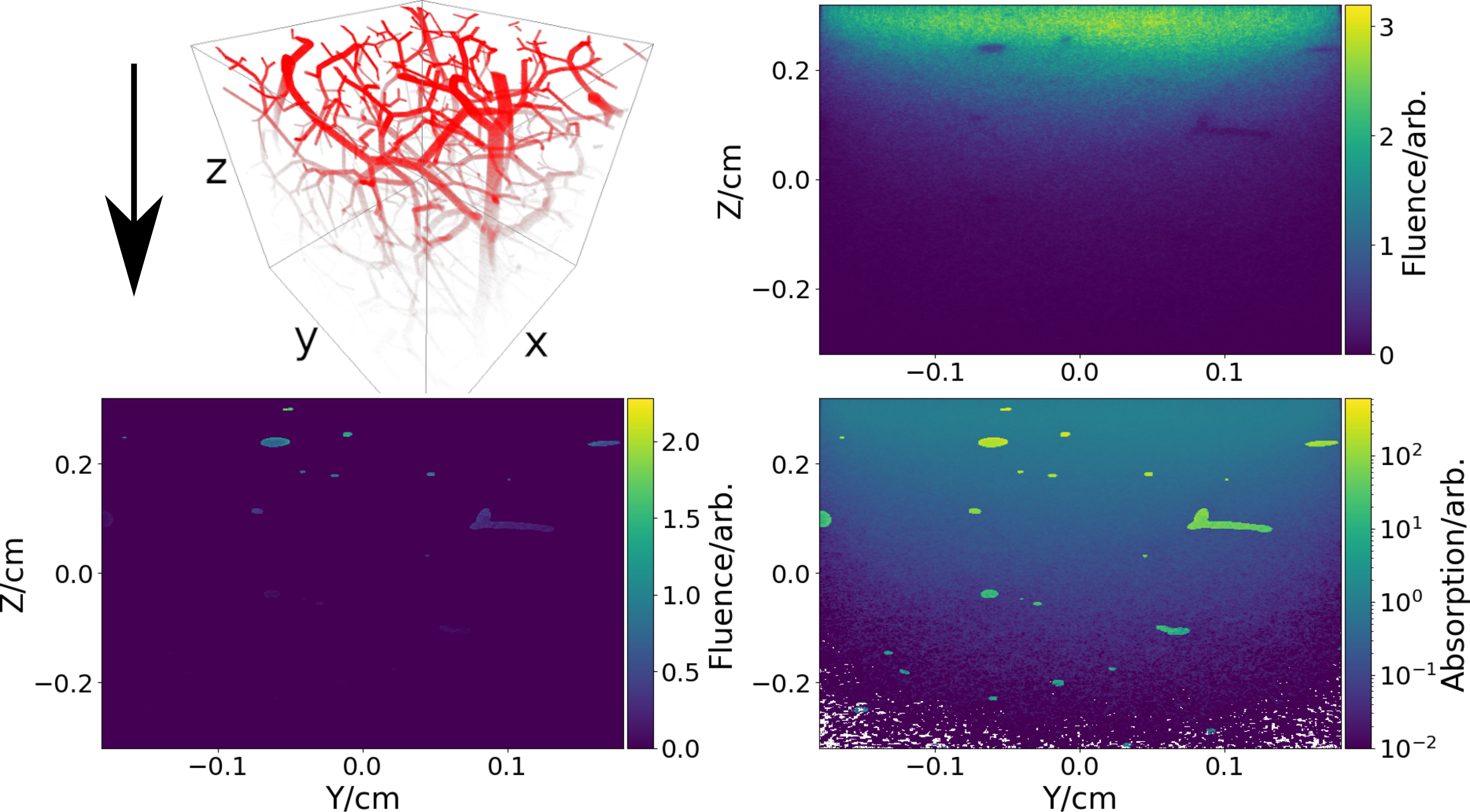}
    \caption{Fluence and absorption images for the vessel network. Light is uniformly incident on the x-y plane of a slab of tissue with an embedded vessel network. Top left panel shows the 3D fluence for the vessel network with tissue's fluence removed for clarity, arrow indicates direction of incident light. Top right shows a slice through the fluence for the tissue and vessels in the y-z plane. Bottom left shows a slice though the fluence for the vessels in the y-z plane. Bottom right shows a slice of absorption for the tissue and vessels on the y-z plane.}
    \label{fig:vessel}
\end{figure}

The vessels are a 3D synthetic microvascular network from data published in~\cite{tetteh2020deepvesselnet} and preprocessed by Yuan \textit{et al.}~\cite{yuan2021light}. 
We model the slab of tissue using a box SDF (second shape in bottom left panel of Figure~\ref{fig:gallery}), and the vessels as a collection of capsule SDFs (third shape in top left panel of Figure~\ref{fig:gallery}) which are then joined together using the CSG operator union (bottom left operation in top right panel of Figure~\ref{fig:gallery}).
The simulation volume is 326 $\times$ 305 $\times$ 611 ~\si{\micro\metre\tothe{3}} and we use a voxel grid of $400^3$ to record the fluence.
The optical properties of the slab of tissue and the vessels are taken from~\cite{marti2018mcmatlab}, and are shown in Table~\ref{tab:optporpves}.
The slab is uniformly illuminated on its top surface by 10 million photons, which are allowed to propagate until they are absorbed or leave the simulated medium.
Figure~\ref{fig:vessel} shows the fluence on the vessel network, and slices of fluence through the tissue slab.

\begin{table}[!ht]
\caption{Table of optical properties for the tissue and vessel network.}
\centering
\begin{tabular}{lllll}
                             & Absorption                      & Scattering                      &                          &      \\
\multicolumn{1}{l|}{}        & \multicolumn{1}{l|}{$\mu_a$ [\si{\per\centi\metre}]} & \multicolumn{1}{l|}{$\mu_s$ [\si{\per\centi\metre}]} & \multicolumn{1}{l|}{g}   & n    \\ \hline
\multicolumn{1}{l|}{Skin}    & \multicolumn{1}{l|}{0.459}      & \multicolumn{1}{l|}{357}        & \multicolumn{1}{l|}{0.9} & 1.38 \\
\multicolumn{1}{l|}{Vessels} & \multicolumn{1}{l|}{231}        & \multicolumn{1}{l|}{94.0}       & \multicolumn{1}{l|}{0.9} & 1.38
\end{tabular}
\label{tab:optporpves}
\end{table}

\section{Conclusion and Outlook}

We have shown a novel meshless, geometrical method for Monte Carlo radiation transport, using signed distance functions.
SDF based models achieve higher accuracy than voxel based models, particularly for modeling smooth surfaces, such as computing fluence in droplets or accurate modeling of human anatomy for light transport calculations.
For problems where the detection of photons is the key measurement and the calculation of light distribution is not, SDFs allow the acceleration of MCRT simulations.

Therefore, SDFs offer an improvement on voxel based models, whilst retaining high accuracy and ease of implementation.
However, there are a number of potential downsides to using SDFs.

In certain configurations the number of steps needed to be taken by a photon packet can be extremely large, see SI Figure 3.
This occurs when the photon is approximately parallel to a surface whilst the distance between the photon and the surface is small.
Recent work has been undertaken to alleviate this problem.
This includes segment tracing which accelerates the sphere tracing method by improving the marching step computation and enhanced sphere tracing which uses an over-relaxation-based method for accelerating sphere tracing~\cite{galin2020segment, korndorfer2014enhanced}.

Large collections of SDFs can also cause massive slow downs due having to evaluate every SDF each time the photon needs to be moved.
This is analogous to the same issue in Monte Carlo models which use triangular meshes.
As in the triangular mesh case, this can be diminished by using a space-partitioning data structure.

The final potential downside is the combination of multiple CSG operations can lead to non-bounded SDFs. Non-bounded SDFs can pose a problem in terms of accuracy and speed~\cite{quilezinterior}. In terms of speed, non-bounded SDFs only give a conservative distance to the surface, resulting in more SDF evaluations which can cause computational slow down. The accuracy problem only affects the computation of surface normals, and is therefore only applicable at refractive index interfaces.

Despite these potential downsides, we envision MCRT codes using SDFs to model the geometry to probe problems such as: effect of skin color on pulse oximtry accuracy, fluence calculation of droplets with viral loads such as Covid-19, and accurate simulations of light propagation in fruit.
In these problems, using SDF over voxels would allow accurate modeling of curved surfaces allowing better accuracy in the simulations.

\section*{Disclosures} Funding: The work was supported by funding from the UK Engineering and Physical Sciences Research Council (EP/P030017/1 and EP/R004854/1) and the H2020 FETOPEN project ``Dynamic'' (EC-GA 863203).

\subsection*{Data, Materials, and Code Availability} 
All code is publicly available at: \url{https://github.com/lewisfish/signedMCRT}.
Data is available at \url{https://doi.org/10.5281/zenodo.5780513}.

\bibliographystyle{unsrt}
\bibliography{references}  


\end{document}


\maketitle

\section{Derivation of Average number of scattering in an isotropic sphere}

For a photon's random walk from the center to the edge of a uniformly scattering sphere of radius $r$.
Consider the net displacement for a single photon from a starting point $p$ after $N$ mean free paths is:
\begin{equation}
    R = r_1+r_2+...+r_N
\end{equation}

The mean square displacement traveled by a photon ($l_*$) is thus:

\begin{align}
    l_*^2=\left<R^2\right>&=\left<r_1^2\right> + \left<r_2^2\right> + ... + \left<r_N^2\right> + 2 \left<r_1 \cdot r_2\right> + ... \\
    \text{where}& \nonumber\\
    2 \left<r_2 \cdot r_2\right> &= 2 \left<|r_1| |r_2| \cos\delta\right> \nonumber
\end{align}

For isotropic scattering $\cos\delta = 0$ therefore all cross terms vanish.
As each term involving the square of the displacement averages to the mean square of the free path of a photon, therefore

\begin{equation}
    l_*^2=N\left<r^2 \right>
\end{equation}
Then multiplying both sides by $\mu^2$, the scattering coefficient, gives an expression for the average number of scatterings for optically thick media.

\begin{equation}
\mu^2 l_*^2 = N \left<\tau^2\right>
\end{equation}

By definition, the scattering coefficient multiplied by the radius of the sphere is $\tau_{max}$:

\begin{equation}
\tau_{max}^2 \equiv N \left<\tau^2\right>
\end{equation}

Then using the following identity
\begin{equation}
\left<\tau^2\right>=\int_0^{\infty} p(\tau)\tau^2d\tau =\int_0^{\infty} e^{-\tau}\tau^2d\tau = 2
\end{equation}
we arrive at an expression for the average number of scatterings as a function of maximum optical depth for an isotropic sphere in the optically thick limit:
\begin{equation}
N = \frac{\tau^2_{max}}{2}
\end{equation}

For the optically thin limit, the number of scatterings is small, on the order of $1-e^{-\tau}\approx\tau$.

\section{Examples of sMCRT on complex geometries}
\begin{figure}[!htbp]
    \centering
    \includegraphics[width=0.75\textwidth]{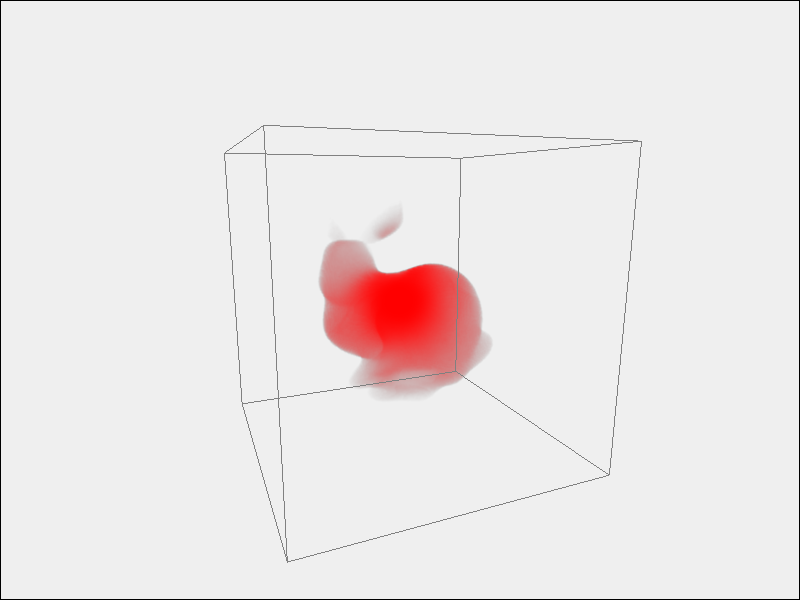}
    \caption{Example of an arbitrary SDF generated via a neural SDF method (SIREN~\cite{sitzmann2020implicit}). Image shows 3D fluence inside the Stanford bunny~\cite{turk1994zippered} for a point source inside the Bunny. Fluence outside the bunny has been removed for clarity.}
    \label{fig:bunny}
\end{figure}

\begin{figure}[!htbp]
    \centering
    \includegraphics[width=0.75\textwidth]{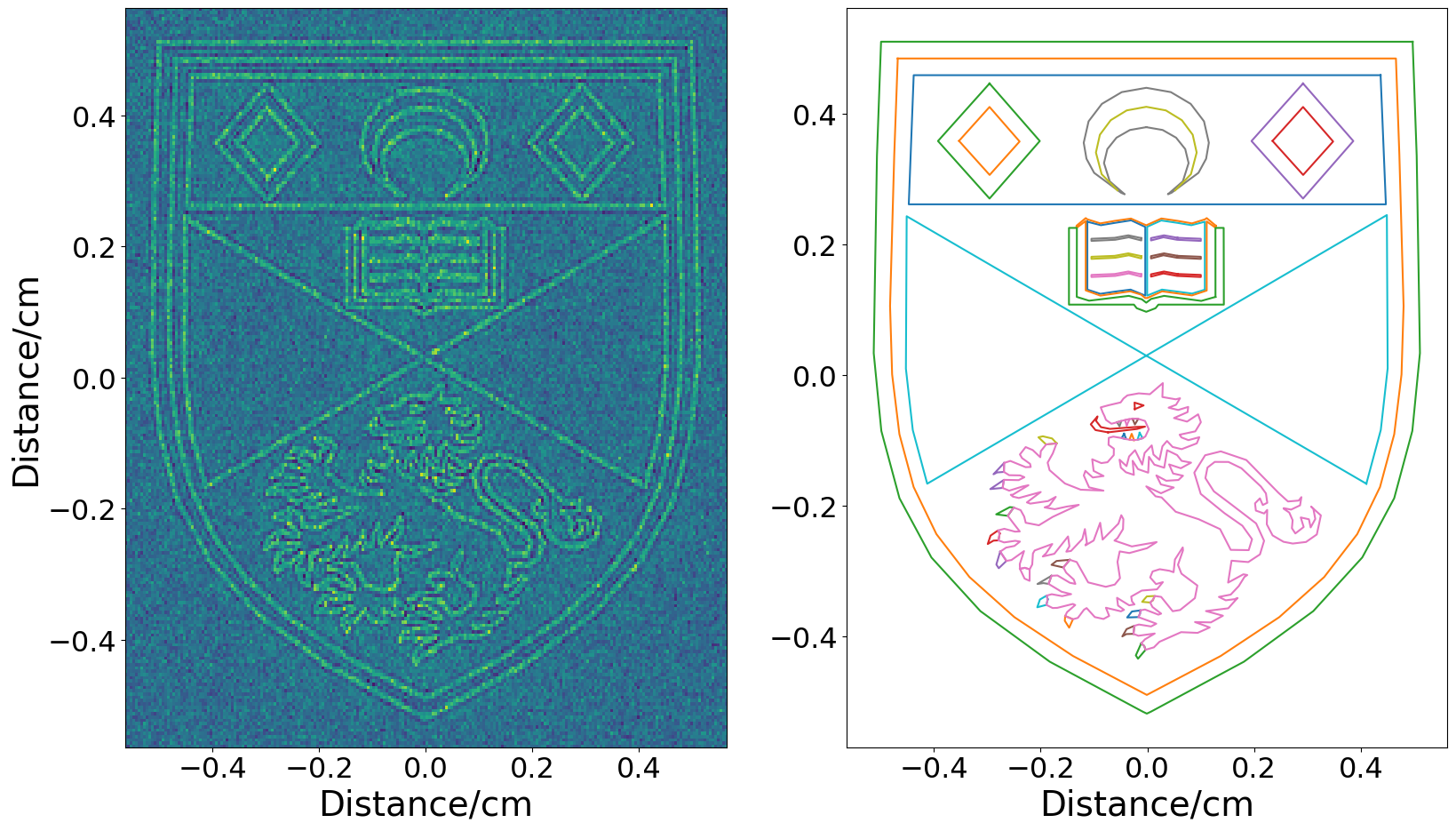}
    \caption{Fluence on the University of St Andrews crest. Crest is converted from a simplified Scalable Vector Graphic (SVG) of the crest, and translated into 2D line segments. Line segments are then extruded in the z-axis and assigned optical properties, where $\mu_a$=0.1~\si{\per\centi\metre}, $\mu_s$=10.0~\si{\per\centi\metre}, g=0.9, and n=1.5. Left shows a slice of fluence in the x-y plane in the middle of the simulated medium. Photons are incident on the crest's surface (into the page). Right shows the simplified SVG where each line has a different color. SVG was simplified by removing some elements, and converting Bézier curves into line segments using Inkscape.}
    \label{fig:crest}
\end{figure}

\clearpage
\section{Number of SDF evaluations}
\begin{figure}[!htbp]
    \centering
    \includegraphics[width=0.75\textwidth]{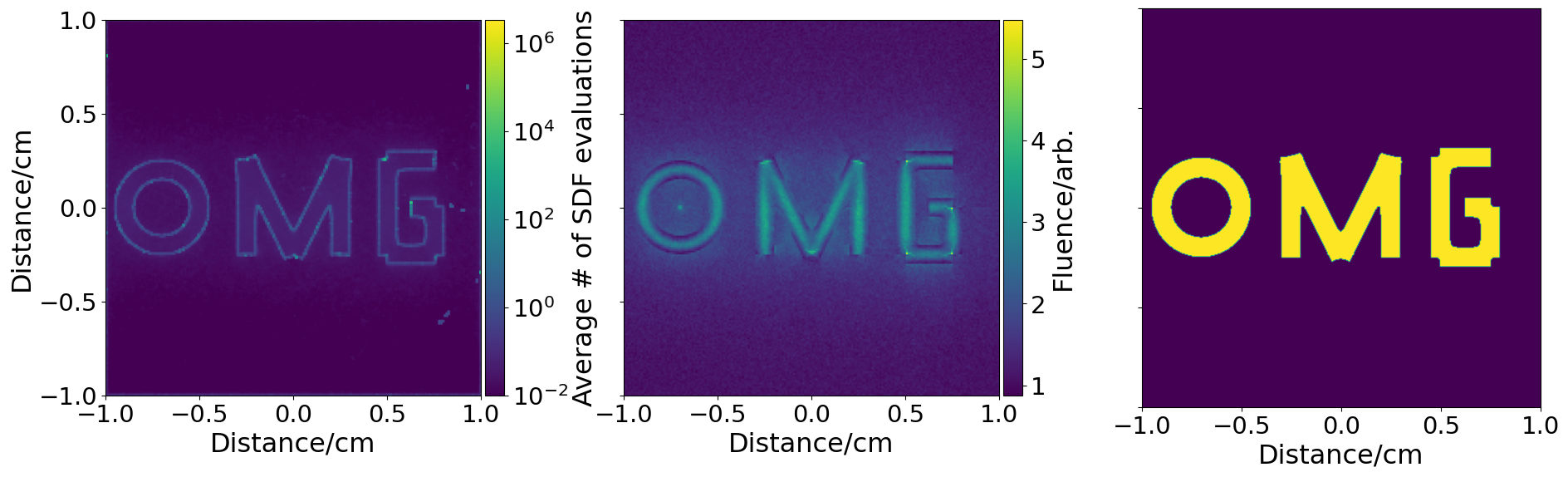}
    \caption{Left panel shows the average number of SDF evaluations for a slice through the middle of the model shown in the right panel. Middle panel shows a slice of fluence through the middle for uniform illumination going into the page.}
    \label{fig:sdfevals}
\end{figure}

\clearpage
\bibliographystyle{unsrt}
\bibliography{references}  